# A Radiation Tolerant Light Pulser for Particle Physics Applications

A. Grummer *, M. R. Hoeferkamp, S. Seidel

Department of Physics and Astronomy, University of New Mexico, MSC 07 4220, 1919 Lomas Blvd. NE, Albuquerque, NM 87131, USA

**ABSTRACT**

A light emitting diode (LED) pulser has been developed that can be used for tests or calibration of timing and amplitude sensitivity of particle physics detectors. A comparative study is performed on the components and pulser output characteristics before and after application of 800 MeV protons and cobalt-60 gammas. This device is demonstrated to be tolerant to fluences up to $6.7 \times 10^{13}$ 800-MeV-p/cm$^2$ and gamma doses up to 5 Mrad.

Keywords: Radiation tolerant, Particle Physics, Detector calibration, Light emitting diode, Proton fluence, Gamma dose

## 1. Introduction

Optical pulsers are frequently needed as in-situ time and amplitude calibrators in particle physics experiments. In applications with no significant radiation field, approaches [1-8, and references therein] to calibration based on radioluminescence sources, optoelectronics with LEDs or lasers, and gas-discharge (lamps) have been employed. A circuit has been developed on the model of Kapustinsky [9] and has been implemented in the design of a pulser for use in radiation fields, for example during commissioning of systems that provide precision timing of particle transit near an interaction point at a hadron collider.

## 2. Design of the Pulser

The circuit, see Fig. 1, uses eight off-the-shelf components plus (at Output) the Thorlabs LED465E, which has a 465 nm wavelength (blue) output. Light pulse intensity is varied by setting the $V_{IN}$ which charges a 100 pF capacitor. Fast discharge of the capacitor through the LED is achieved with a low impedance transistor switch (complementary pair of RF bipolar junction transistors) which is triggered by an externally generated trigger pulse. The circuit was implemented in pairs on laminated glass epoxy boards with finished thickness 0.06", dimensions 1.5" × 1.95", surface mount components, and 1-oz. copper trace layers.

## 3. Characterization

Pulsers were characterized before and after exposure to a proton beam or to cobalt-60 gammas. Signals from a photomultiplier tube (PMT) were read by a Tektronix TDS7354B 2.5 GHz oscilloscope. Fig. 2 shows the experimental setup. The distance from LED to PMT is 60 cm. The PMT used for measurements before and after gamma irradiation is a Hamamatsu R7400U. The PMT used for measurements before and after proton irradiation is a Photonis 85001-501. The effects of the R7400U PMT and the TDS7354B oscilloscope on the width of the signal from the pulser range from 0.3% at $V_{IN}$ = 4 V to 0.03% at $V_{IN}$ = 24 V; these values are calculated using the transient time spread (TTS) of the PMT (230 ps) and the effect of the oscilloscope bandwidth (176 ps). The same values were assumed for the Photonis 85001-501 PMT. The standard deviation of the pulse width of a typical non-irradiated LED ranges from 6.2% at $V_{IN}$ = 4 V to 7% at $V_{IN}$ = 24 V. In Fig. 3 we show a typical measurement of a normalized pulse width at two different values of $V_{IN}$. This measurement technique is sufficient for comparative measurements. An absolute timing measurement [10] would require single photon counting using a constant fraction discriminator (CFD).

## 4. Proton irradiations

Three pulser boards, including all their components, were exposed to 800 MeV protons at the Los Alamos LANSCE facility in 2013. The protons were applied at room temperature over a period of less than an hour. Two boards received $1.5 \times 10^{13}$ p/cm$^2$ and one received $6.7 \times 10^{13}$ p/cm$^2$. Fig. 4 indicates the effect of the proton fluences upon the pulse width, as a function of input voltage. The error bars shown reflect variations of the $V_{IN}$ (0.5%), angle and distance of the LED with respect

---

* Corresponding author:
*Email Address:* agrummer@unm.edu (A. Grummer)

Jan. 25, 2018

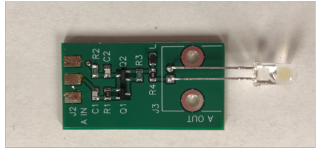
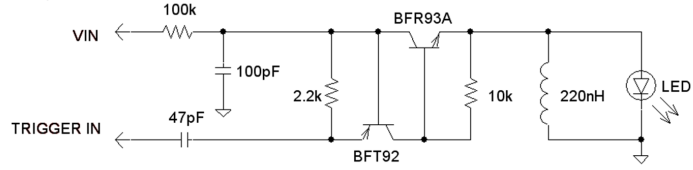

**Fig. 1.** The pulser circuit and board.

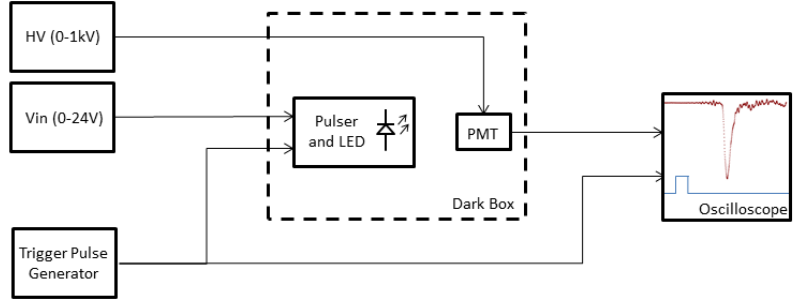

**Fig. 2.** The measurement setup.

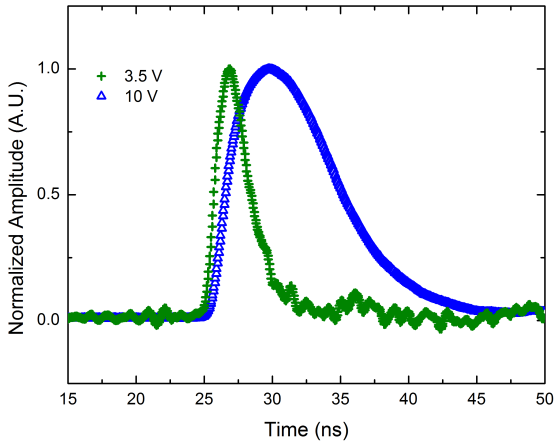

**Fig. 3.** Output pulse width of a non-irradiated pulser board at $V_{IN}$ = 3.5 V and $V_{IN}$ = 10 V.

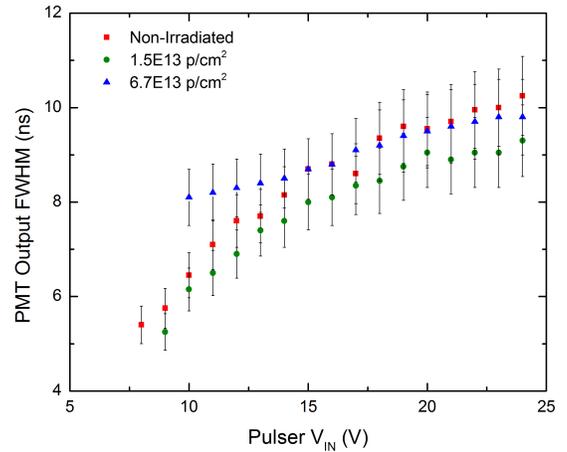

**Fig. 4.** Pulse width of the pulser before and after proton irradiation to fluences of $1.5 \times 10^{13}$ and $6.7 \times 10^{13}$ 800-MeV-p/cm$^2$.

to the PMT (0.03%), and oscilloscope precision (1%), as well as the PMT TTS and oscilloscope bandwidth. Also included in the error is the statistical variation in the typical pulse width of an LED465E previously mentioned. After irradiation with $1.5 \times 10^{13}$ p/cm$^2$, the pulse shape and minimum width are similar to the pre-irradiated values. The minimum pulse width rises from 5 to 8 ns after application of $6.7 \times 10^{13}$ p/cm$^2$. The minimum $V_{IN}$ at which signal output is achievable increases with dose. Pulse width increases as $V_{IN}$ increases in each set of data. Fig. 5 shows the pulse amplitude ($V_{OUT}$) as a function of input voltage for the same devices. Contributing factors to the error are variations of the $V_{IN}$ (ranging from 12% at $V_{IN}$ = 7 V to 0.3% at $V_{IN}$ = 24 V), angle and distance of the LED with respect to the PMT (0.1%), and oscilloscope precision (1%). Although the amplitude is diminished in the proton-irradiated devices, over 50% of the light output is retained in the upper half of the input voltage range.

## 5. Gamma Irradiations

Ten pulsers were exposed to gammas at the Sandia Gamma Irradiation Facility in 2017, with two each receiving doses of 100 krad, 500 krad, 1 Mrad, 3 Mrad, and 5 Mrad. Fig. 6 shows the output pulse amplitude as a function of input voltage, before and after irradiation for all doses. The LED light diminishes incrementally with increasing gamma dose, reaching a minimum of 80% for the 5 Mrad case. Fig. 7 shows the output pulse width as a function of input voltage, before and after irradiation for all doses. The pulse width increases monotonically with increasing $V_{IN}$ and the gamma dose applied does not affect the FWHM significantly. The error bars shown are calculated in the same way as mentioned previously. The pulser circuit is robust to this gamma dose.

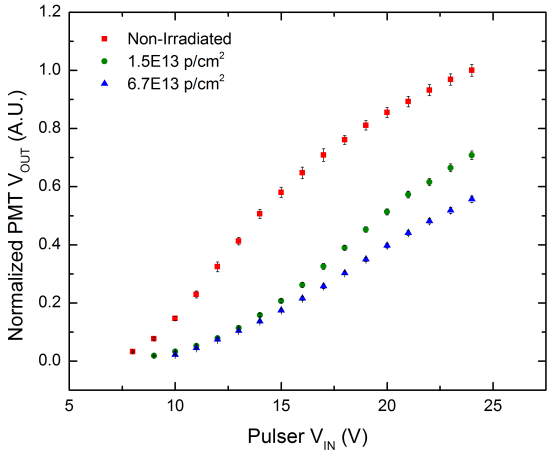

**Fig. 5.** Pulse amplitude of the pulser before and after proton irradiation to fluences of $1.5 \times 10^{13}$ and $6.7 \times 10^{13}$ 800-MeV-p/cm$^2$.

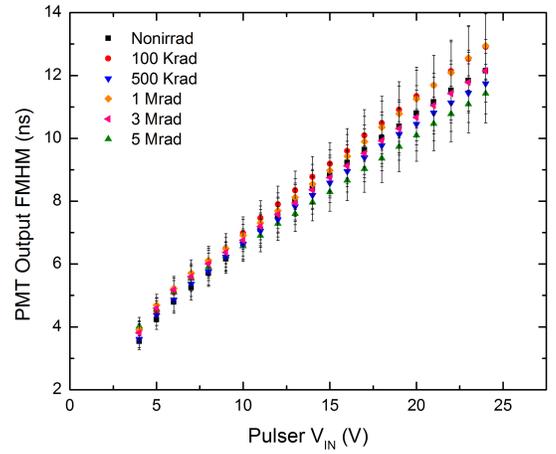

**Fig. 7.** Pulse width of the pulser before and after gamma irradiation to doses of 0.1, 0.5, 1.0, 3.0, and 5.0 Mrad. Data shown are the averaged values of two devices at each dose.

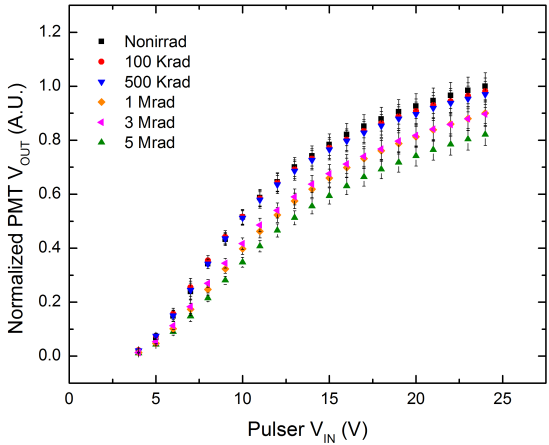

**Fig. 6.** Pulse amplitude of the pulser before and after gamma irradiation to doses of 0.1, 0.5, 1.0, 3.0, and 5.0 Mrad. Data shown are the averaged values of two devices at each dose.

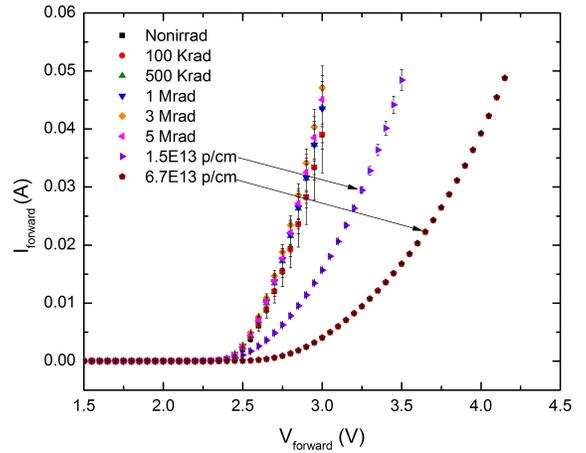

**Fig. 8.** Measured forward current and applied forward voltage of irradiated versus non-irradiated LED 465E components.

## 6. Discussion

The gamma and proton irradiated components were removed from the pulser boards after irradiation. The

LEDs' current versus voltage (IV) characteristics were measured (see Fig. 8). The IV curves of the gamma irradiated LEDs overlap the IV curves of the non-irradiated ones. In the proton irradiated devices the forward threshold voltage is increased and for any given forward voltage the output current (and thus the output light) is diminished in proportion to the applied proton fluence. Fig. 9 examines the effect of the radiation on the transistors' gain; we show here an example of the $I_C$ vs $V_{BE}$ of the BFR93A NPN transistor. Data from the gamma-irradiated devices again overlap the data from the non-irradiated transistors. The proton irradiated transistors exhibit a higher shift in $V_{BE}$ threshold voltage and a decrease in transistor gain. Similar results are measured on the BFT92 PNP transistor. All other components (thick film resistors, ceramic capacitors, and a wirewound ferrite core inductor) were found to have been unaffected by both the gamma and the proton exposures. Visual inspection shows the LED to be discolored by both gamma and proton exposure. We conclude that the reduction of the PMT output amplitude of the gamma irradiated devices is due to discoloration of the clear LED casing. The reduction of the PMT output amplitude seen in the proton irradiated devices is due to the effects on the semiconductor material in the LEDs and transistors as well as the LED casing discoloration.

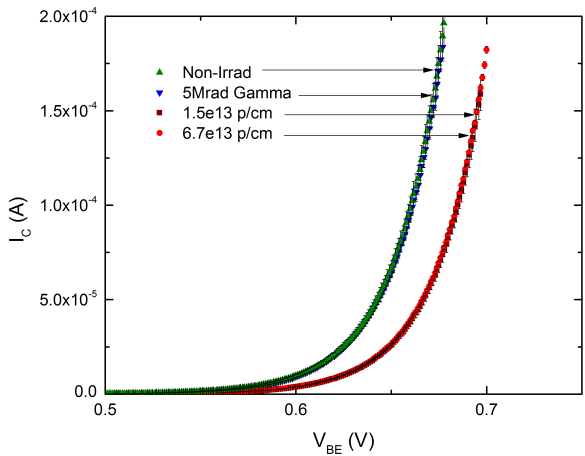

**Fig. 9.** Measured $I_C$ and applied $V_{BE}$ of irradiated versus non-irradiated NPN transistors BFR93A at the constant value of $V_{CE} = 2V$.

## 7. Conclusion

A pulser circuit suitable for operation in radiation fields up to at least $6.7 \times 10^{13}$ 800-MeV-protons/cm$^2$ and 5 Mrad gammas has been developed. Pulse widths are largely unchanged due to the irradiations and remain in the 4-12 ns range depending upon input voltage. Output amplitude and thus the light output decreases proportionally to the proton and gamma exposure but preserves 80% (in the case of gammas) and 50% (in the case of hadrons) at the highest exposure. The affected components of the pulser board are the RF transistors and the LEDs. The LED cases are sensitive to both species, while the semiconductors are changed by the hadrons only. The device is compact, economical, and applicable to a variety of particle physics experimental environments.

## Acknowledgements

We thank the instrument scientists and accelerator operators at the LANSCE facility, especially Dr. Ron Nelson, for technical support and the excellent operation of the accelerator. We thank the technical staff at the Sandia GIF, especially Dr. Maryla Wasiolek, for very productive collaboration and technical assistance. We thank Haley McDuff for assisting with some data collection. This work is supported by U.S. Department of Energy grant DE-SC0018006.